# Spin-orbit torque generation in bilayers composed of CoFeB and epitaxial SrIrO$_3$ grown on an orthorhombic DyScO$_3$ substrate

Sosuke Hori[1], Kohei Ueda*[1,2], Takanori Kida[3], Masayuki Hagiwara[3] and Jobu Matsuno[1,2]

[1]*Department of Physics, Graduate School of Science, Osaka University, Osaka 560-0043, Japan*
[2]*Center for Spintronics Research Network, Graduate School of Engineering Science, Osaka University, Osaka 560-8531, Japan*
[3]*Center for Advanced High Magnetic Field Science, Graduate School of Science, Osaka University, Osaka 560-0043, Japan*

**ABSTRACT**

We report on the highly-efficient spin-orbit torque (SOT) generation in epitaxial SrIrO$_3$ (SIO) which is grown on an orthorhombic DyScO$_3$(110) substrate. By conducting harmonic Hall measurement in Co$_{20}$Fe$_{60}$B$_{20}$ (CoFeB)/SIO bilayers, we characterize two kinds of the SOTs, i.e., dampinglike (DL) and fieldlike ones to find that the former is much larger than the latter. By comparison with the Pt control sample with the same CoFeB thickness, the observed DL SOT efficiency $\xi_{DL}$ of SIO (~0.32) is three times higher than that of Pt (~0.093). The $\xi_{DL}$ is nearly constant as a function of the CoFeB thickness, suggesting that the SIO plays a crucial role in the large SOT generation. These results on the CoFeB/SIO bilayers highlight that the epitaxial SIO is promising for low-current and reliable spin-orbit torque-controlled devices.

*kueda@phys.sci.osaka-u.ac.jp

Current-driven spin-orbit torque (SOT) in ferromagnet (FM)/non-magnet (NM) bilayers has been attracting plenty of attention due to its possibility of device application with low-power-consumption driving technology. By applying a charge current to NM with strong spin-orbit coupling (SOC), a spin current is induced in perpendicular to the charge current, resulting from the bulk spin Hall effect (SHE)[1]. At the interface of the bilayers, the spin current via the SHE gives rise to a spin accumulation, which causes the SOT acting on the magnetization in FM layer[2]. Since the SOT is a powerful way to switch the magnetization[3–6], it is desirable to enhance the efficiency of SOT generation. While much effort so far has been done on 5$d$ transition metals with large SHE, e.g., Pt[4,7–15], Ta[3,7,9,14,16–19], and W[5,20], exploring other material systems that exhibit highly-efficient SOT generation constitutes an important challenge towards spintronic application.

Recently, 5$d$ transition-metal oxides have emerged as a class of spintronic materials due to their unique electronic structure dominated by 5$d$ electrons with strong SOC[21]. The 5$d$ oxides contrast well with the 5$d$ transition metals of which the electronic structure is dominated by 6$s$ electrons as well as 5$d$ electrons. Ir oxides out of the 5$d$ oxides become a potential platform as spintronic compounds; several groups have reported SOT generation in amorphous $IrO_2$[22,23], epitaxial $IrO_2$[24], and epitaxial $SrIrO_3$ (SIO)[25–29]. Among them, particularly interesting is the epitaxial SIO[25–29]. Epitaxial strain induced by substrate plays a crucial role in stabilizing SIO with the perovskite structure, whereas the bulk perovskite SIO[30] is synthesized only under a high-pressure condition. The epitaxial strain also allows control of the crystal symmetry and hence the electronic structure of SIO; it strongly affects the SOT generation through SOC[31,32]. Substrate choice could be therefore one of the factors for highly-efficient SOT generation, while this is not the case with non-epitaxial $IrO_2$[22,23] or 5$d$ transition metals[3–20]. In the previous research on the SOT generation in SIO, $SrTiO_3$ (STO)[25,27–29] and $(LaAlO_3)_{0.3}(SrAl_{0.5}Ta_{0.5}O_3)_{0.7}$ (LSAT)[26] substrates have been used so far. We here focus on $DyScO_3$ (DSO) as another choice of substrate in order to clarify a spintronic characteristic of bulk SIO; the lattice constants of DSO are much closer to those of SIO than those of STO and LSAT substrates. Moreover, an orthorhombic structure of DSO enables us to deposit orthorhombic SIO films. A prior study[25] has shown the importance of orthorhombic SIO, which generates the larger SOT compared with that of tetragonal one. Investigating the SIO grown on the DSO is therefore expected to provide further experimental insight into relationship between the crystal structure and the SOT generation.

In this Letter, we report on highly-efficient SOT generation in SIO epitaxially grown on DSO substrate that minimizes the strain effect, in contrast to the above-mentioned substrates. We estimate the SOT with dampinglike (DL) and fieldlike (FL) symmetries in $Co_{20}Fe_{60}B_{20}$ (CoFeB)/SIO bilayers and those in Pt control sample, by performing harmonic Hall measurement. The observed DL SOT in SIO is more than three times of that in Pt. In both SIO and Pt, the FL SOT is about the one-third of the DL SOT. The DL SOT efficiency $\xi_{DL}$ is independent of the CoFeB thickness, revealing negligible spin current generation from the CoFeB layer. These results emphasize that the strong SOC of the 5$d$ electrons plays an important role in the large SOT generation in the epitaxial SIO.

The SIO films were grown on DSO substrate by a pulsed laser deposition from ceramic SIO target using a KrF excimer laser ($\lambda$ = 248 nm) at 5 Hz. In the bulk form, SIO and DSO exhibit an orthorhombic GdFeO$_3$ type perovskite structure, of which the pseudocubic lattice parameter $a_{pc}$ are ~0.3950 nm[30] and ~0.3940 nm[33], respectively. This lattice mismatch of ~0.3% is small compared to that of ~0.9% for STO and ~2.1% for LSAT substrates; we can expect that coherent epitaxial growth of SIO on the DSO(110)$_O$ substrate is easily realized. Substrate temperature and oxygen pressure during the deposition were 650 ℃ and 25 Pa, respectively. The crystalline quality and thickness of the SIO films were confirmed by x-ray diffraction measurement. Figure 1(a) shows the 2$\theta$-$\theta$ scan in the SIO grown on DSO(110)$_O$ substrate. The film and substrate peaks are represented by the pseudocubic and orthorhombic indexed as (hkl)$_C$ and (hkl)$_O$, respectively. We find two significant peaks (001)$_C$ and (002)$_C$ of the SIO film corresponding to DSO orientations without any impurity peaks, indicating that single-phase SIO film is epitaxially grown. Figure 1(b) is the magnified view of Fig. 1(a), showing the peaks of the SIO(002)c film and the DSO(110)o substrate. The SIO(002)$_C$ peak gives the out-of-plane lattice constant of 0.3955 nm, which is in good agreement with reported values[33–35]. The clear oscillation stemming from thickness evidences a good crystallinity of the film, which is also supported by narrow full width at half maximum of the rocking curve around the SIO(002)c (~0.05 deg.). From the oscillation, we determined the film thickness to be 21 nm. Figure 1(c) exemplifies the reciprocal space mapping, which represents the horizontal axis and vertical axis corresponding to the SIO(010)$_C$ [or DSO(001)$_O$] and SIO(001)$_C$ [or DSO(110)$_O$], respectively. We also conducted the reciprocal space mapping containing the SIO(100)$_C$ axis [or the axis perpendicular to the DSO(110)$_O$ and (001)$_O$]. These results clearly show that the film is coherently strained, indicating the SIO grown on the DSO substrate has orthorhombic structure. Hence, all the diffraction data illustrates that the high-quality orthorhombic SIO film is epitaxially grown on the DSO(110)o substrate.

In order to evaluate SOT in SIO, we prepare bilayers consisting of epitaxial SIO and ferromagnetic CoFeB. The whole film structures are represented by TaO$_x$(2.5)/CoFeB($t_{CoFeB}$)/SIO(21), where the number in parenthesis indicates the thickness in nanometer [Fig. 2(a)]. After the SIO growth, Ta and CoFeB were deposited *ex situ* by a radio-frequency magnetron sputtering at Ar deposition pressure of 0.3 Pa. The TaO$_x$ layer is obtained from the as-deposited Ta metal by natural oxidation in the air; it prevents the CoFeB layer from being oxidized. We also confirmed that no SOT contribution from the TaO$_x$ layer is found by performing control experiment. The CoFeB thickness $t_{CoFeB}$ ranges from 2 to 6 nm. The CoFeB and TaO$_x$ thickness were estimated from growth rate of each layer determined by x-ray reflectivity measurement beforehand. Considering that Pt has been known to generate large SOT, we also prepare a Pt control sample *in situ*, that is, TaO$_x$(2.5)/CoFeB(2)/Pt(5) on thermally oxidized Si substrate. The CoFeB/SIO and CoFeB/Pt bilayers were fabricated into a Hall bar with two arms by photolithography and Ar ion milling. Ta(5)/Pt(60) contact pads were attached at the end of devices for electrical measurement. The Hall bar has channel dimensions of 30 μm length ($L$) and 10 μm width ($w$), as shown in Fig. 2(a). The $\phi$ represents the azimuthal angles of the external magnetic field ($B_{\text{ext}}$). We apply an ac current $I_{\text{ac}} = \sqrt{2} I_{\text{rms}} \sin(2\pi f t)$ with root mean square of current $I_{\text{rms}}$ in the $x$ axis direction and frequency $f$ = 13 Hz. We set $I_{\text{rms}}$ to be 0.4 mA for longitudinal resistance ($R$) measurement in $x$-axis direction, and 1 mA for Hall resistance ($R_{\text{H}}$) measurement and 1.5 mA for harmonic Hall measurement in the $y$-axis directions. The resistivity ($\rho$) of SIO and CoFeB are 570 and 190 μΩcm, respectively, obtained from the linear fit on the $t_{\text{CoFeB}}$ versus the inverse sheet $R$ as displayed in Fig. 2(b); the fitting procedure was performed only for CoFeB(3,4,5)/SIO(21) with common SIO thickness since the SIO thickness slightly changes with the bilayer films, i.e., 23 nm for CoFeB(2), 21 nm for CoFeB(3,4,5), and 20 nm for CoFeB(6). The obtained resistivities of the CoFeB and SIO layers well reproduce the resistance for CoFeB(2)/SIO(23) and CoFeB(6)/SIO(20) with different SIO thickness, suggesting the consistency of the whole resistance data.

We evaluate the SOT by conducting harmonic Hall measurement[7,9,13–16,18,19–23,26–29]. By applying the $I_{\text{ac}}$ to the Hall bar, the SOT is induced at FM/NM interface, which has two components with different symmetries, namely, DL and FL SOTs. These SOTs correspond to DL effective field $B_{\text{DL}} \parallel (\sigma \times m)$ and FL effective field $B_{\text{FL}} \parallel \sigma$, where $m$ and $\sigma$ is the direction of the magnetization and accumulated spin polarization; the latter is along the $y$-axis direction. Of these two SOTs, the DL SOT is most relevant to the magnetization switching[3–6]. In order to distinguish the $B_{\text{DL}}$ and $B_{\text{FL}}$, the harmonic Hall measurements were performed by measuring an angle dependence of first and second harmonic resistance ($R_{\text{H}}^{1\omega}$, $R_{\text{H}}^{2\omega}$) at fixed $B_{\text{ext}}$ in $xy$ plane; the applied $B_{\text{ext}}$ varies 0.1 to 1.2 T. The $R_{\text{H}}^{1\omega}$ corresponds to the conventional dc Hall measurement, while the $R_{\text{H}}^{2\omega}$ reflects the influence of SOT. These obey the following Eq. (1), (2)[7,14,19,22,23,26–29].

$$R_{\text{H}}^{1\omega} = R_{\text{PHE}} \sin 2\phi \quad (1)$$

$$R_{\text{H}}^{2\omega} = -\left(R_{\text{AHE}} \frac{B_{\text{DL}}}{B_{\text{ext}} + B_{\text{k}}} + R_{\nabla T}\right) \cos \phi + 2 R_{\text{PHE}} \frac{B_{\text{FL}} + B_{\text{Oe}}}{B_{\text{ext}}} (2\cos^3 \phi - \cos \phi) \quad (2)$$

$$\equiv -R_{\text{DL}+\nabla T} \cos \phi + R_{\text{FL+Oe}} (2\cos^3 \phi - \cos \phi) \# \quad (3)$$

Here, $R_{\text{PHE}}$, $R_{\text{AHE}}$, $B_{\text{k}}$, $R_{\nabla T}$, and $B_{\text{Oe}}$ correspond to planar Hall resistance, anomalous Hall resistance, out-of-plane anisotropy field, thermal induced second-harmonic resistance driven by a temperature gradient, e.g. the anomalous Nernst effect[36] and spin Seebeck effect[37], and current induced Oersted field, respectively. $R_{\text{AHE}}$ and $B_{\text{k}}$ were estimated by Hall resistance depending on the out-of-plane $B_{\text{ext}}$. The upper panel of Fig. 2(c) shows the $R_{\text{H}}^{1\omega}$ as a function of $\phi$. The $R_{\text{H}}^{1\omega}$ slightly deviates from Eq. (1), which is attributed to anomalous Hall effect (AHE) resulting from a small sample misalignment[7]. Since we confirmed that this AHE contribution has negligible effects on estimate of $R_{\text{PHE}}$, we determined the $R_{\text{PHE}}$ from Eq. (1). The $\phi$ dependence of $R_{\text{H}}^{2\omega}$ is induced by the small modulation of the magnetization from its equilibrium position due to the current-driven SOTs. We can define the $B_{\text{DL}}$ contribution ($R_{\text{DL}+\nabla T}$) and the $B_{\text{FL}}$ contribution ($R_{\text{FL+Oe}}$) as the coefficients of the $\cos\phi$ and $(2\cos^3\phi - \cos\phi)$ in Eq. (3). The bottom panel of the Fig.2 (c) shows the $R_H^{2\omega}$ of CoFeB(6)/SIO(21) measured at 0.1 and 1 T, which is well fitted by Eq. (2). The difference in amplitude of the $R_H^{2\omega}$ depending on the $B_{ext}$ indicates the suppression of SOTs by the $B_{ext}$. We estimated the $R_{DL+\nabla T}$ and $R_{FL+Oe}$ from the $R_H^{2\omega}$ fitting, which is displayed in the top panel and the bottom panel in Fig. 2(d), respectively. Figure 2(d) indicates the large $R_{DL+\nabla T}$ contribution and the small $R_{FL+Oe}$ contribution to the $R_H^{2\omega}$ in CoFeB/SIO bilayers.

Then we extract two SOT effective fields $B_{DL}$ and $B_{FL}$ through the coefficients $R_{DL+\nabla T}$ and $R_{FL+Oe}$. Figure 3(a) and 3(b) show the $R_{DL+\nabla T}$ for CoFeB(2)/SIO(21) and CoFeB(2)/Pt(5). The data indicates linear dependence on $1/(B_{ext} + B_k)$ in accordance with Eq. (3). The data points in the low-field region deviate from the linear fitting, possibly due to unsaturated in-plane magnetization. The slopes and intercepts of the $R_{DL+\nabla T}$ correspond to the $B_{DL}$ and the $R_{\nabla T}$, respectively. We estimated the DL effective field per current density $B_{DL}/J$, where $J$ is the applied charge current density flowing in the SIO (Pt) layer. The estimated $B_{DL}/J$ is +4.3 mT/($10^{11}$Am$^{-2}$) for CoFeB(2)/SIO and +1.1 mT/($10^{11}$Am$^{-2}$) for CoFeB(2)/Pt, suggesting the sizable DL SOT generation in CoFeB/SIO as well as in CoFeB/Pt. The $R_{\nabla T}$ of CoFeB/SIO is much larger than CoFeB/Pt. The temperature gradient in CoFeB/SIO is larger compared with CoFeB/Pt due to the higher resistivity of SIO than that of Pt, resulting in the large $R_{\nabla T}$. The $R_{FL+Oe}$ for CoFeB(2)/SIO(21) and CoFeB(2)/Pt(5) are plotted as a function of $1/B_{ext}$ as shown in Fig. 3(c) and 3(d). The $R_{FL+Oe}$ for CoFeB(2)/Pt(5) has linear dependence on $1/B_{ext}$, which is well explained by Eq. (3), while we assume the $(B_{FL} + B_{Oe}) \sim 0$ for CoFeB(2)/SIO(21) since the linear fit is impractical because of its large error. The $B_{FL}$ were obtained by subtracting the $B_{Oe}$ contribution estimated by Ampere's law as $B_{Oe} = \mu_0 J d_{SIO}/2$, where $\mu_0$ is the magnetic permeability in a vacuum. The $B_{FL}/J$ for CoFeB(2)/Pt and CoFeB(2)/SIO are $+4.1 \times 10^{-1}$ and +1.4 mT/($10^{11}$Am$^{-2}$), respectively, similar to previous values in CoFeB/Pt bilayer[15] and Py/SIO bilayers[26] performed by harmonic Hall measurement. In the following, we only focus on the $B_{DL}$ since the FL component in CoFeB/SIO includes the experimental uncertainty in its absolute value.

In order to further discuss the SOT, we evaluate the $\xi_{DL}$ from the $B_{DL}/J$ using following Eq. (4)[38]:

$$\xi_{DL} = \frac{2e\mu_0 M_s t_{CoFeB}}{\hbar} \frac{B_{DL}}{J} \qquad (4)$$

where $e$, $M_s$, and $\hbar$ are the elementary charge, the saturation magnetization, and the Dirac constant, respectively. The $M_s$ for CoFeB/SIO and CoFeB/Pt are $1.1 \times 10^6$ and $1.3 \times 10^6$ A/m, respectively, measured by the superconducting quantum interference device magnetometer, in agreement with those for typical CoFeB thin film[3,14,16,17,19]. We obtain $\xi_{DL} = +0.32$ for SIO and +0.093 for Pt at $t_{CoFeB} = 2$ nm as shown in Fig. 4, showing that the $\xi_{DL}$ for SIO is around three times larger than that for Pt. Note that our result of $\xi_{DL} = +0.093$ for Pt is consistent with the reported values[4,7,8,10,11,13,15], demonstrating validity of our experimental setup. We also consider that the $\xi_{DL}$ is saturated due to the large NM layer thickness sufficiently above the spin-diffusion length, i.e., ~1.7 nm for SIO[25] and ~1.4 nm for Pt[11]. Here, we refer that recent studies point out spin current generation stemming from the AHE and/or SHE in FM layer[39–42]. In order to clarify the issue, the $\xi_{DL}$ as a function of the $t_{CoFeB}$ is displayed in Fig. 4. The $\xi_{DL}$ shows no appreciable dependence on the $t_{CoFeB}$, suggesting negligible spin current contribution in the CoFeB layer; this evidence highlights that the obtained $\xi_{DL}$ are truly derived from the SIO layer. Note that we confirmed that the $\xi_{DL}$ is independent of current direction relative to crystallographic directions; this might be attributed to the small in-plane crystalline anisotropy of the SIO due to the DSO substrate with a tiny in-plane anisotropy (~0.1 %).

As found in the previous studies, the $\xi_{DL}$ of the SIO ranges from +0.2 to +1.1 for the thick SIO layers[25–29]; these studies are performed in bilayers with different FM layers such as metals and oxides, indicating that the interfacial condition would influence the estimation of the $\xi_{DL}$. With the factor in mind, our result is consistent with the previous studies. The observed large $\xi_{DL}$ of +0.32 for SIO on DSO substrate reinforces the validity of SIO as the spintronic material considering that the lattice constants of our SIO films are very close to those of bulk SIO. This large SOT generation is also compatible with the importance of the orthorhombicity of the SIO reported in the previous study[25]. Our findings would be an important step to clarify the charge to spin current conversion in epitaxial $5d$ oxide thin films, which show a variety of intriguing physical properties: Dirac nodal line[43], spin-orbit driven semimetal-Mott insulator transition[44,45], and non-linear planar Hall effect[46]. By incorporating these properties, further studies on spintronics using SIO are motivated.

In conclusion, we studied the SOT generation in CoFeB/SIO bilayers, where the SIO layer is epitaxially grown on DSO substrate. By performing the harmonic Hall measurement in CoFeB/SIO bilayers, we estimated that the $\xi_{DL}$ of the SIO is around three times larger than that of the Pt control sample. Further investigation shows little dependence of the CoFeB thickness on the $\xi_{DL}$ (0.32–0.36) and hence negligible contribution of the SOT from the CoFeB layer. Our findings indicate highly-efficient SOT generation in the SIO layer, which may be utilized for magnetic tunnel junction device driven by SOT switching based on $5d$ transition metal oxides.

## ACKNOWLEDGEMENT

The authors thank T. Arakawa for technical support. This work was supported by Nanotechnology Platform of MEXT, Grant Number JPMXP09S21OS0027. This work was also partly supported by the JSPS KAKENHI (Grant Nos. JP19K1543, JP19H05823, and JP20H05160), JPMJCR1901 (JST-CREST), and Nippon Sheet Glass foundation for Materials Science and Engineering. We acknowledge stimulating discussions at the meeting of the Cooperative Research Project of the Research Institute of Electrical Communication, Tohoku

University

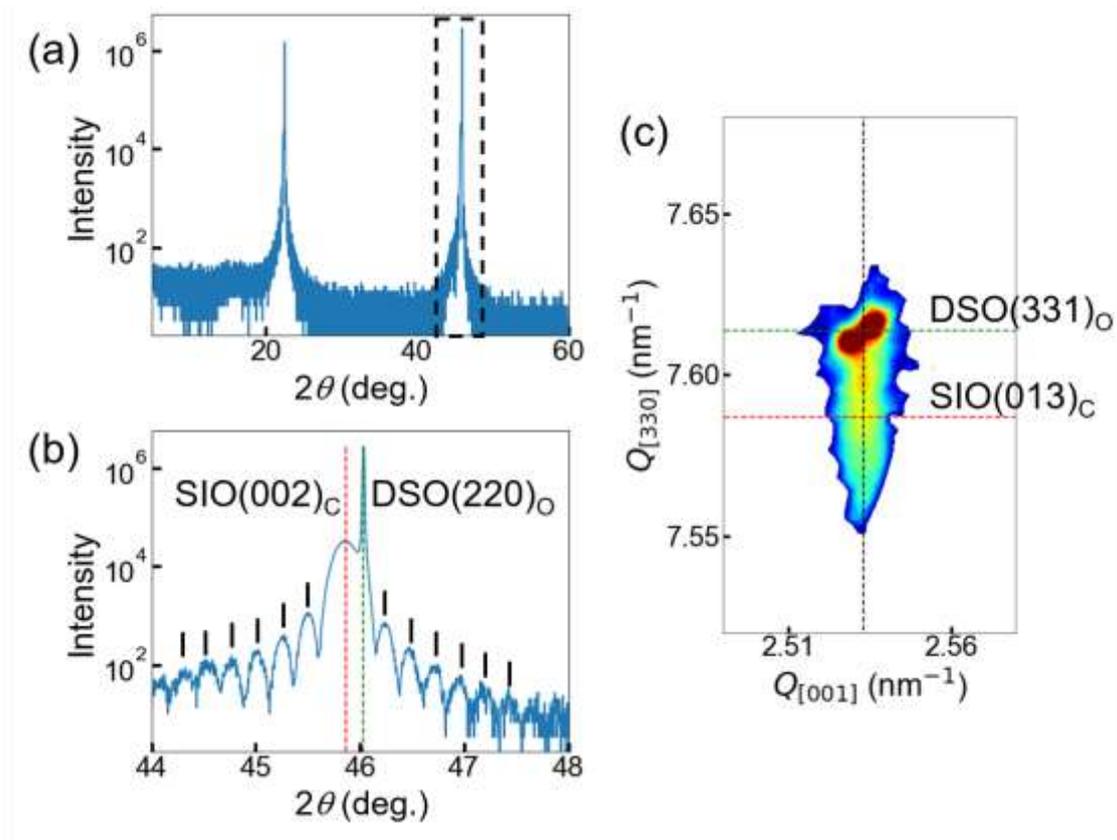

**Fig. 1.** (a) X-ray diffraction $2\theta\text{-}\theta$ scan of a SrIrO$_3$ (SIO) film grown on a (110)$_O$-orientated DyScO$_3$ (DSO) substrate. Peaks of the SIO(001)$_C$ and (002)$_C$ are located in the vicinity of the corresponding DSO(110)$_O$ and (220)$_O$ peaks. Note that the film and substrate peaks are indexed as cubic (002)$_C$ and orthorhombic (220)$_O$, respectively. (b) Magnified view of the scan around SIO(002)$_C$ reflection in (a). (c) Reciprocal space mapping of (001)$_C$ oriented SIO film epitaxially grown on the corresponding DSO orientation.

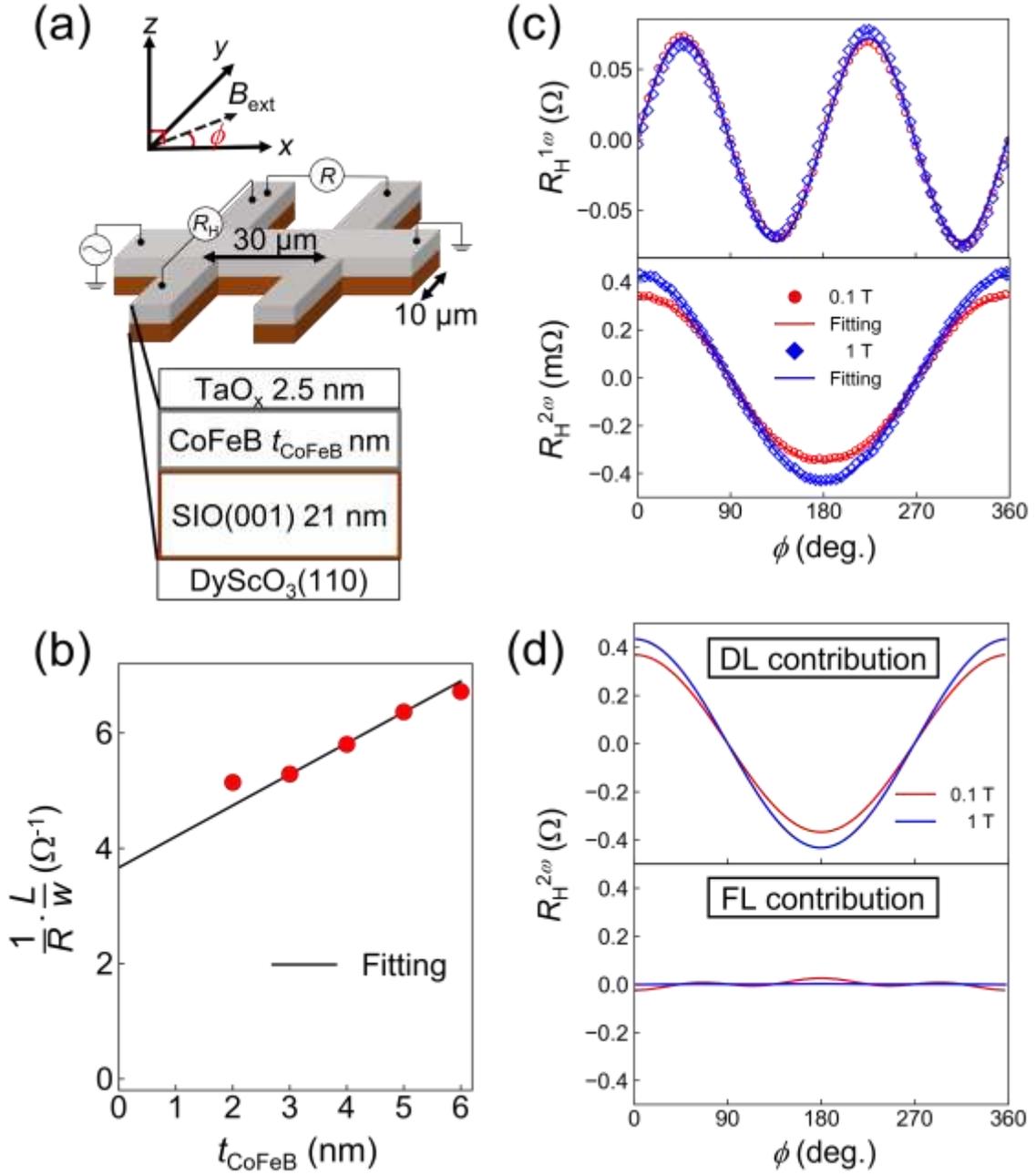

**Fig. 2.** (a) Schematic illustration of the device and cross section of the samples. The $\phi$ represent the azimuthal angles of the external magnetic field ($B_{ext}$). AC current is applied along the $x$ axis direction to detect Hall resistance ($R_H$) in the $y$ axis direction. (b) The inverse sheet resistance $1/R \cdot L/w$ as a function of $t_{CoFeB}$. The sold line represents a linear fit for $t_{CoFeB}$ = 2, 3, and 4 nm. (c) $R_H^{1\omega}$ and $R_H^{2\omega}$ of CoFeB(6)/SIO(21) measured at 0.1 and 1 T. The solid curves of $R_H^{1\omega}$ and $R_H^{2\omega}$ are fits to the data using Eq. (1, 2). (d) $R_H^{2\omega}$ are separated to $\cos\phi$ and ($2\cos^3\phi - \cos\phi$) components from the fittings, which indicate DL contribution and FL contribution, respectively.

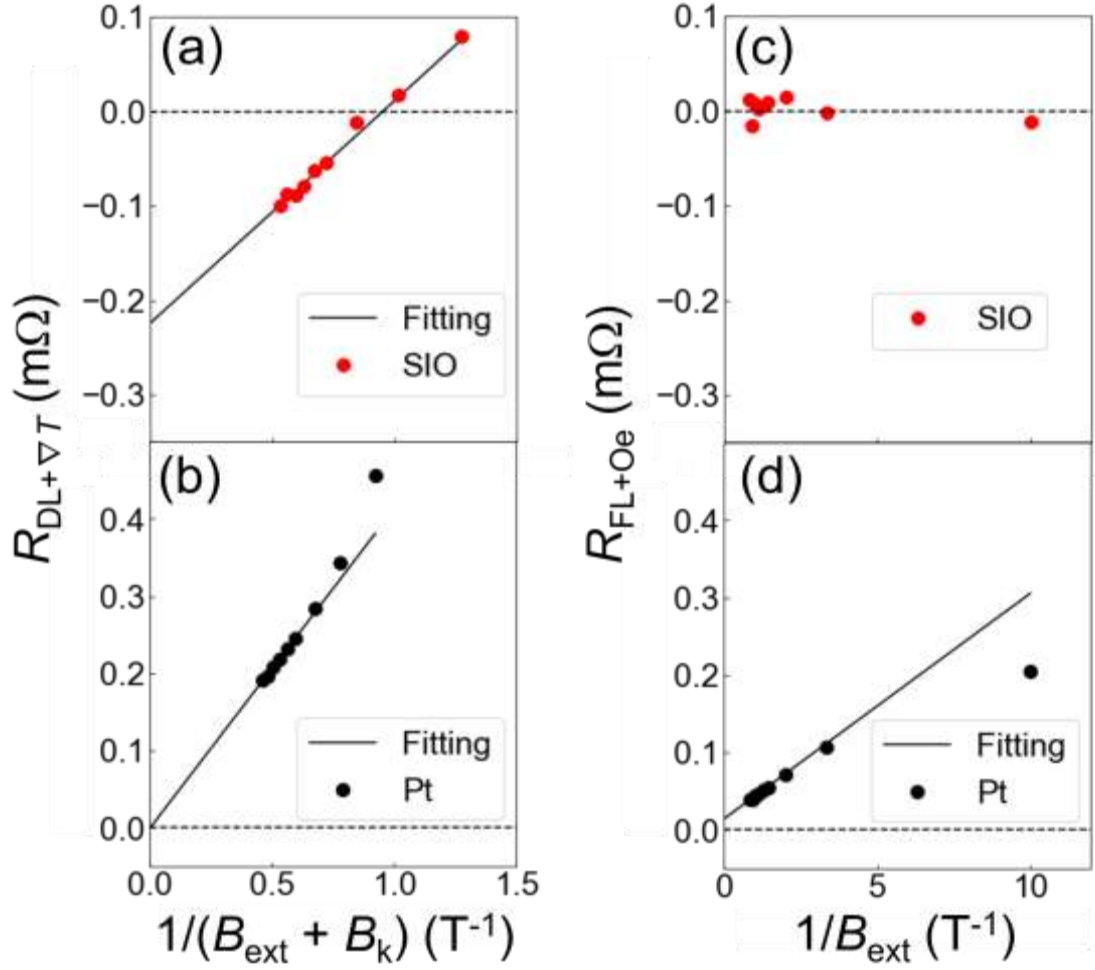

**Fig. 3.** (a), (b) $R_{DL+\nabla T}$ as a function of $1/(B_{ext} + B_k)$ in CoFeB(2)/SIO and CoFeB(2)/Pt(5) samples. (c), (d) $R_{FL+Oe}$ as a function of $1/B_{ext}$ in CoFeB(2)/SIO(21) and CoFeB(2)/Pt(5) samples. The solid lines are linear fits to the experimental data.

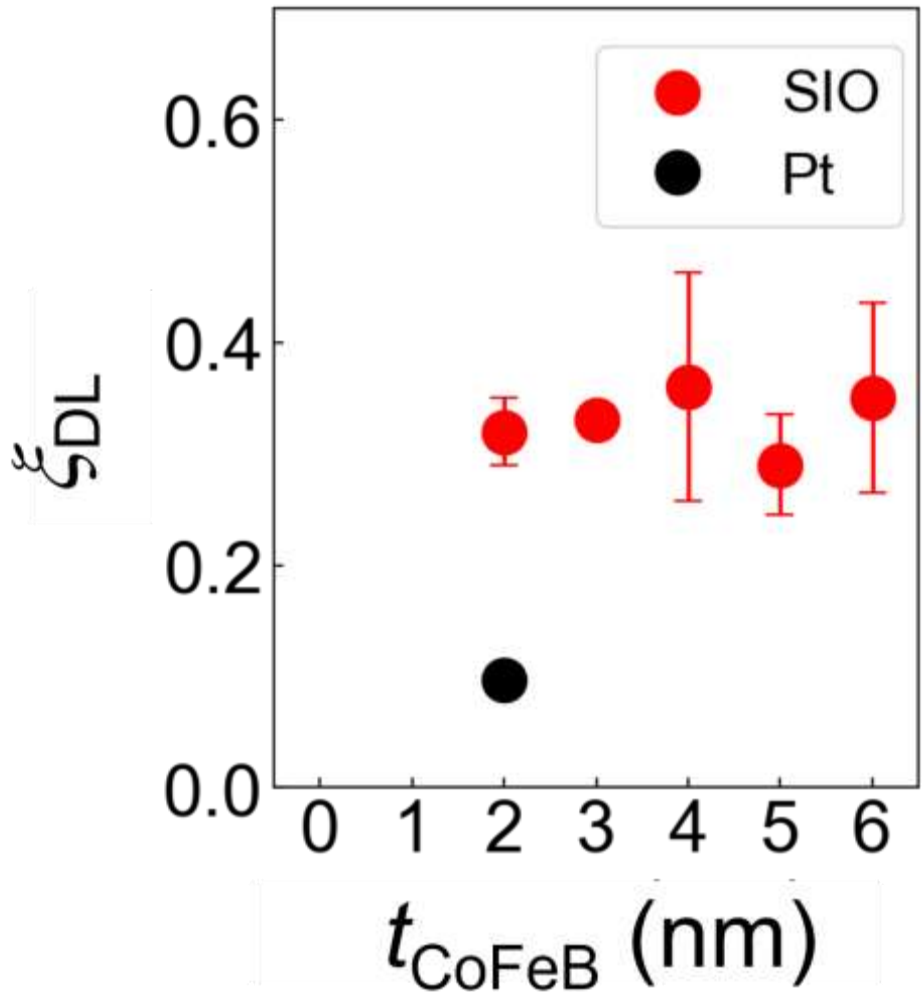

**Fig. 4.** DL SOT efficiency $\xi_{DL}$ as a function of the CoFeB thickness, and of Pt control sample.